\documentclass[11pt,showpacs,preprintnumbers,amsmath,amssymb,prd,nofootinbib,superscriptaddress]{revtex4-2}

\usepackage{dcolumn}% Align table columns on decimal point
\usepackage{bm}% bold math
\usepackage{ifpdf}
\usepackage{hyperref}
\usepackage{xcolor,color,graphicx,graphics,physics}
\usepackage[spanish,english]{babel}%, portuguese
\usepackage[latin1]{inputenc}
\usepackage[OT1]{fontenc}
\usepackage{latexsym,amssymb,amsmath,amsfonts, slashed,cancel}
\usepackage{makeidx}
\usepackage{epsfig,subfigure}
\usepackage{natbib}
\usepackage{epstopdf}
\usepackage{mathrsfs}
\usepackage{hyperref}%\usepackage[colorlinks=true,linkcolor=blue]{hyperref}
\hypersetup{colorlinks=true, linkcolor=blue, citecolor=blue, urlcolor=blue}
\usepackage{enumerate}

\usepackage{fixmath}

\everymath{\displaystyle}
\usepackage{graphicx}

\usepackage[T1]{fontenc}
\usepackage{amsmath}
\usepackage{amssymb}
\usepackage{graphicx}
\usepackage{xcolor}

\newcommand{\bea}{\begin{eqnarray}}
\newcommand{\eea}{\end{eqnarray}}

\newcommand{\orcid}[1]{\href{https://orcid.org/#1}{\includegraphics[width=10pt]{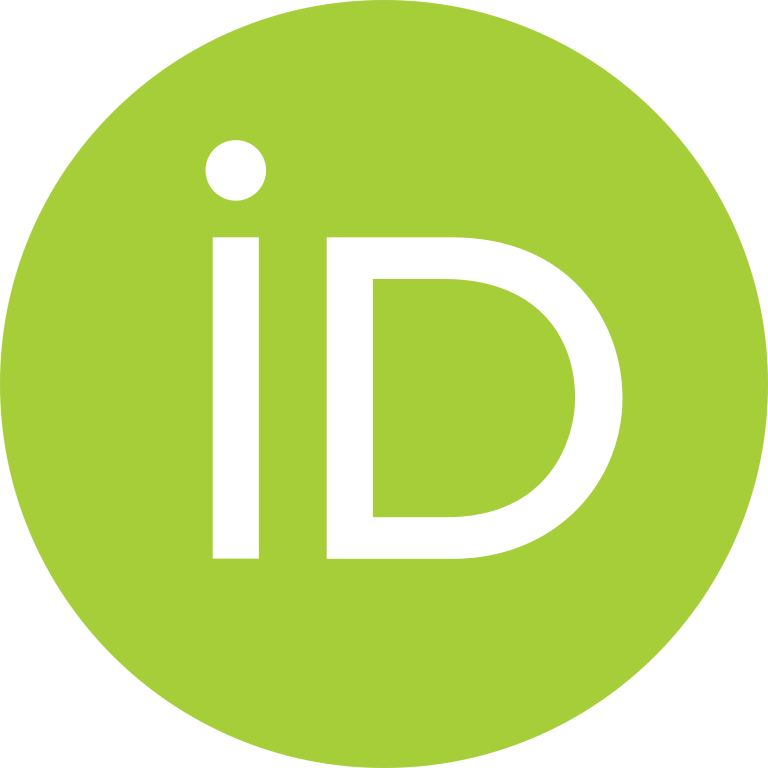}}}

\begin{document}

\title{$e^{+}e^{-}\to l^{+}l^{-}$ scattering at finite temperature in the presence of a classical background magnetic field}

\author{D. S. Cabral  \orcid{0000-0002-7086-5582}}
\email{danielcabral@fisica.ufmt.br}
\affiliation{Instituto de F\'{\i}sica, Universidade Federal de Mato Grosso,\\
78060-900, Cuiab\'{a}, Mato Grosso, Brazil}

\author{A. F. Santos \orcid{0000-0002-2505-5273}}
\email{alesandroferreira@fisica.ufmt.br}
\affiliation{Instituto de F\'{\i}sica, Universidade Federal de Mato Grosso,\\
78060-900, Cuiab\'{a}, Mato Grosso, Brazil}

\begin{abstract}

In this work the $e^{+}e^{-}\to l^{+}l^{-}$ scattering process is investigated. The cross-section is calculated considering three different effects: temperature, external magnetic field and chemical potential.  The effect due to an external field is inserted into the problem through a redefinition of the fermionic field operator. Effects due to temperature and chemical potential are introduced using the Thermo Field Dynamics formalism.

\end{abstract}

\maketitle

\section{Introduction}

The study of processes in quantum field theory at finite temperatures is very important. It serves not only to determine the well-known quantities in a more real situation, adding the thermal effects, but also to obtain constraints and limits that show a description of the primordial world, which are systems under conditions similar to those at the beginning of the universe. In addition to temperature, strong magnetic fields and finite chemical potential play highly important roles, particularly in phase transition regimes \cite{thermalelecpos, dominik}. In order to examine these conditions and features a particular scattering process is considered, namely, electron-positron annihilation decaying in a lepton-antilepton pair, $e^{+}e^{-}\to l^{+}l^{-}$ scattering. 

To investigate the influence of an external background field on the problem, it is added to the quantum photonic field in the QED Lagrangian. The procedure here is similar to that used by \cite{dissertmag}, however, it is adapted to its thermal approach by the Thermo Field Dynamics (TFD) formalism. This, in turn, is a thermal field theory that consists of doubling the Hilbert space by making a copy of it and introducing the temperature-dependent states and operators by a rotation of this doubled space \cite{khannatfd, lietfd}.  There are many works studying processes at finite temperature by different approaches \cite{QEDefetive,comptonfinite}. There are also some investigations using the TFD formalism \cite{daniel1, daniel2, ale1, praseletron, bhabhatfd, khannatfd}. However, despite its importance, in most cases, the contribution of the chemical potential has not been considered.

The connection with thermodynamics, however, is made through the grand canonical partition function instead of the canonical one, which, in addition to the temperature, also introduces the chemical potential into the problem. The  presence of the chemical potential ($\mu$) is very important in a scattering process since it is the energy required to add a particle to a system without changing entropy and volume \cite{howardlee}. Then, the contribution of this term will be considered and calculated. An alternative way to put the chemical potential in the system, which not will be done here, is to change the Lagrangian \cite{persson}.

For the investigation carried out in this work, it is important to observe that the classical background field leads to a modification in the wave function for the fermionic field, as well as in the energy expression. Strictly speaking, this feature modifies the particle and antiparticle spinors by adding to them a new coordinate dependency and a new quantum number that emerges. Therefore, for the first there is a new completeness relation depending on the special Landau functions, and the second gives the possible energies that the system can occupy as a function of the field strength.

This paper is organized as following. In Sec. \ref{secDirac}, it is presented how to describe the Dirac field focusing on the new forms of the energy, as well as on the spinors and their modified completeness relations and, consequently, introducing new terms called Landau functions. Then, to work with the thermal effects of this theory, the formalism and the structure of TFD are presented. In addition, the differential cross-section  is calculated for a $a+b\to c+d$ reaction, looking at the dependence on Landau levels $n$. In Sec. \ref{secpotential}, a review of the meaning of chemical potential in a physical system is made. Its functional dependence on temperature and lepton density is obtained by two different methods: the first is a Sommerfeld approximation and the second is a Taylor series expansion. Then, the interpretation of these results in a QED scattering process and the role of this quantity in a quantum field theory are discussed. Finally, in Sec. \ref{secscattering} the thermal $e^{+}e^{-}\to l^{+}l^{-}$ scattering in the presence of an external classical background magnetic field is investigated, showing the calculations of its cross-section in the limit of very strong fields. Some conclusions remarks are presented in the Sec. \ref{secconclusion}.

In this work,  natural units of particle physics,  $\hbar=c=1$, have been used. The scattering diagrams are made with the Jaxodraw application \cite{jaxo}. The calculations are performed with Mathematica software \cite{mathematica}.

\section{The Dirac field}\label{secDirac}

Let us consider a charged massive fermion field $\psi$ whose charge and mass are $q$ and $M$, respectively. The Hamiltonian describing this particle in the presence of a magnetic field is written as
\begin{eqnarray}
H=\vec{\alpha}\cdot\left(-i\vec{\nabla}-q\vec{A}^{(B)}\right)+\mathbb{T} M,\label{eq05}
\end{eqnarray}
where $\vec{A}^{(B)}$ is the background classical potential vector \cite{kazama} and
\begin{eqnarray}
\vec{\alpha}=\begin{pmatrix}
0 & \vec{\sigma}\\\vec{\sigma} & 0
\end{pmatrix},\quad\quad\quad \mathbb{T}=\begin{pmatrix}
I & 0 \\ 0 & -I
\end{pmatrix},
\end{eqnarray}
with $\vec{\sigma}$ and $I$ being the Pauli and identity $2\times2$ matrices, respectively.

In this way, the Lagrangian of the system is given as
\begin{eqnarray}
\mathcal{L}=\bar{\psi}\left(i\gamma^\nu\Pi_\nu-M\right)\psi,  \label{eq04}
\end{eqnarray}
where  $\gamma^\nu$ are the Dirac matrices and $\Pi_\nu=\partial_\nu+iqA_\nu^{(B)}$, which looks like a covariant derivative, but with the background field playing the role of the gauge field. From Eq. (\ref{eq05}) the motion equations, i.e. the modified Dirac equation, can be extracted, which becomes the eigenvalue relation $H\psi=E\psi$. Then the equation to be solved is
\begin{eqnarray}
i\gamma^\nu\Pi_\nu\psi=M\psi,\label{eq12}
\end{eqnarray} 
whose solutions are given by
\begin{eqnarray}
\psi=e^{-iEt}\begin{pmatrix}
\psi_R(\vec{r})\\\psi_L(\vec{r})
\end{pmatrix},\label{eq65}
\end{eqnarray}
with $(t,\vec{r})=(t,x,y,z)$ being the position in space-time. The spinorial field $\psi$ is a column vector represented by $\psi_R$ and $\psi_L$.

It is important to note that, choosing a constant magnetic field in the z-direction, the potential is given by
\begin{eqnarray}
A^{\nu}_{(B)}=\left(0,-yB,0,0\right),\label{eq21}
\end{eqnarray}
such that, in Cartesian coordinates, $\vec{B}=\vec{\nabla}\times \vec{A}=B\hat{z}$, with $B>0$ being the amplitude of the field.

Therefore, under these conditions, taking $q=-e$ as the electron charge, the form of Eq. (\ref{eq65}) in addition to Eq. (\ref{eq12}) gives us a differential equation whose solutions are 
\begin{eqnarray}
\psi_{R,L}(x^\nu)=C_{R,L}^{(1)}u_{R,L}e^{-ip_\nu x^\nu_{\cancel{y}}}+C_{R,L}^{(2)}v_{R,L}e^{ip_\nu x^\nu_{\cancel{y}}},\label{eq66}
\end{eqnarray}
where the notation $x^{\nu}_{\cancel{y}}$ is adopted and represents a space-time position without $y$-component, this implies $p_\nu x^{\nu}_{\cancel{y}}=p_0t-p_xx-p_zz$. The indices $R$ and $L$ represent the right- and left-handed spinor solutions while $C^{(j)}_{R,L}$ are constants with $j=1,2$ representing positive and negative energy eigenstates, respectively.

Note in Eq. (\ref{eq66}) that the plane wave parts of the solution have the $y$-component equal to zero, such that this dependence is shown in the $u_s$ and $v_s$ spinors. Doing this, the problem to be solved becomes a special Hermite problem \cite{diracmag} and for the positive energy solutions it has the following
\begin{eqnarray}
u_{1}(y,n,\vec{p}_{\cancel{y}})=\begin{pmatrix}
I_{n-1}(\xi^{+})\\0\\\frac{p_z}{E_n+M}I_{n-1}(\xi^{+})\\
-\frac{\sqrt{2neB}}{E_n+M}I_n(\xi^{+})
\end{pmatrix},\quad\quad\quad u_{2}(y,n,\vec{p}_{\cancel{y}})=\begin{pmatrix}
0\\I_n(\xi^{+})\\-\frac{\sqrt{2neB}}{E_n+M}I_{n-1}(\xi^{+})\\-\frac{p_z}{E_n+M}I_n(\xi^{+})
\end{pmatrix},\label{eq67}
\end{eqnarray}
and for the negative energy solutions,
\begin{eqnarray}
v_{1}(y,n,\vec{p}_{\cancel{y}})=\begin{pmatrix}
\frac{p_z}{E_n+M}I_{n-1}(\xi^{-})\\ \frac{\sqrt{2neB}}{E_n+M}I_{n}(\xi^{-})\\I_{n-1}(\xi^{-})\\0
\end{pmatrix},\quad\quad\quad v_{2}(y,n,\vec{p}_{\cancel{y}})=\begin{pmatrix}
\frac{\sqrt{2neB}}{E_n+M}I_{n-1}(\xi^{-})\\ -\frac{p_z}{E_n+M}I_n(\xi^{-})\\0\\I_n(\xi^{-})
\end{pmatrix}.\label{eq68}
\end{eqnarray}

The $s=1,2$ index represents the spin-up and spin-down states and 
\begin{eqnarray}
E_n=\pm\sqrt{M^2+p_z^2+2neB},\label{eq14}
\end{eqnarray}
is the energy spectrum, obtained by the dispersion relation coming from the Hermite equation, with $n=0,1,2,...$. 

The term $I_n$ appearing in the expressions (\ref{eq67}) and (\ref{eq68}) is the $n$-th Landau function given by
\begin{eqnarray}
I_n(\xi)=\left(\frac{\sqrt{eB}}{2^nn!\sqrt{\pi}}\right)^{1/2}e^{-\xi^2/2}H_n(\xi)
\end{eqnarray}
with $H_n(\xi)$ being the Hermite polynomials of order $n$. The functions $I_{n-1}$ are related to spin-up solutions,  while $I_n$ performs the same for spin-down solutions. Furthermore, for $n=0$ there is no solution for the spin-up state, then $I_{-1}\equiv0$. In addition, these functions obey the following completeness relations
\begin{eqnarray}
\sum_n I_n(\xi)I_n(\xi^\prime)=\sqrt{eB}\delta(\xi-\xi^\prime)=\delta(y-y^\prime)
\end{eqnarray}
and
\begin{eqnarray}
\int_{-\infty}^{+\infty}I_n(\xi)I_j(\xi)d\xi=\sqrt{eB}\delta_{n,j}.
\end{eqnarray}

The variable $\xi$ present in the spinor solutions is given by
\begin{eqnarray}
\xi^{\pm}=\sqrt{eB}\left(y\mp\frac{p_x}{eB}\right),
\end{eqnarray}
which is a term that makes a connection between $y$-coordinate and $p_x$-momentum.

Consequently, the fermionic field is written as
\begin{eqnarray}
\psi(x^\nu)=\int\frac{dp_xdp_z}{(2\pi)^2}\sum_{s,n}\sqrt{\frac{E_n+M}{2E_n}}\left[a_s(n,\vec{p}_{\cancel{y}})u_s(y,n,\vec{p}_{\cancel{y}})e^{-ip_\nu x^\nu_{\cancel{y}}}+b_s^{\dagger}(n,\vec{p}_{\cancel{y}})v_s(y,n,\vec{p}_{\cancel{y}})e^{ip_\nu x^\nu_{\cancel{y}}}\right],\label{eq07}
\end{eqnarray}
whose anti-commutation relations are
\begin{eqnarray}
\left\{\psi(x_\nu),\psi^\dagger(x^{\prime}_\nu)\right\}&=&\delta^3(\vec{x}-\vec{x^\prime}),\\
\left\{a_s(n,\vec{p}_{\cancel{y}}),a^\dagger_{s^\prime}(n^\prime,\vec{p^\prime}_{\cancel{y}})\right\}&=&(2\pi)^2\delta_{s,s^\prime}\delta_{n,n^\prime}\delta(p_x-p^\prime_x)\delta(p_z-p^\prime_z),\label{eq13}
\end{eqnarray}
being the same for antiparticle $b$'s operators and equal to zero for the others. The spinors obey the following modified completeness relations
\begin{eqnarray}
\sum_s u_s(y,n,\vec{p}_{\cancel{y}})\bar{u}_s(y^\prime,n,\vec{p}_{\cancel{y}})=\frac{1}{E_n+M}F_u(y,y',n,\vec{p}_{\cancel{y}}),\label{eq27}
\end{eqnarray}
for the positive energy part and
\begin{eqnarray}
\sum_s v_s(y,n,\vec{p}_{\cancel{y}})\bar{v}_s(y^\prime,n,\vec{p}_{\cancel{y}})=\frac{1}{E_n+M}F_v(y,y',n,\vec{p}_{\cancel{y}}),\label{eq28}
\end{eqnarray}
for the negative energy part. Here $F_u$ and $F_v$ are somewhat complicated functions depending on Landau terms $I_n$ \cite{dissertmag,betadecaymag}. 

In addition, due to the $y$-dependence, the spinors also satisfy the following continuous properties
\begin{eqnarray}
\int_{-\infty}^{\infty}dy\left[u^{\dagger}_s(y,n,p_{\cancel{y}})u_{s^{\prime}}(y,n^\prime,p_{\cancel{y}})\right]=\int_{-\infty}^{\infty}dy\left[v^{\dagger}_s(y,n,p_{\cancel{y}})v_{s^{\prime}}(y,n^\prime,p_{\cancel{y}})\right]=\delta_{s,s^\prime}\delta_{n,n^\prime}\frac{2E_n}{E_n+M},\label{eq30}
\end{eqnarray}
and the other combinations as $u^\dagger v$ and $v^\dagger u$ are equal to zero.

Some notes about Eq. (\ref{eq07}) should be considered. (i) The field operator is a superposition of plane waves, and the exponentials guarantee this fact, but it is necessary to note that the $p_y$ component does not appear on the expression because the $y$ component of space-time vanishes. (ii)  The spinors do not depend on $p_y$ since they explicitly depend on Landau functions of the space. In other words, the $y$-component of momentum is not zero, but undefined by the uncertainly principle, and therefore the integral cannot be evaluated in this direction because it will lead to divergence.

Finally, these $n$ indices of Landau functions, whose energy is strongly dependent, although very large, do not go to infinity. This can be seen from the spectra (\ref{eq14}). For a fermionic incident particle with mass $M$ and whose energy $E$ is given, from the dispersion relation one can extract the limit $E^2-M^2\geq 2neB$. In other words, if $2neB>E^2-M^2$ it would be necessary for $p_z$ to be a complex number, which is impossible because it is a measurable physical quantity \cite{diracmag}.

Up to now, it has been described how to treat fermions in the presence of a classical magnetic field. Where the new operator given by Eq. (\ref{eq07}) and its adjoint $\bar{\psi}=\psi^\dagger\gamma^0$ with the anti-commutation rule Eq. (\ref{eq13}) configures all Dirac mechanics, in this special case, within a background potential, but still at zero temperature.

To make this field theory thermal, the Thermo Field Dynamics (TFD) formalism can be used. This approach is based on the concepts of Lie algebra and symmetries, in such a manner that it is characterized by doubling the Hilbert space \cite{lietfd,khannatfd}. In other words, in this formalism, the thermal effects are added to the theory through the thermal space $\mathbb{H}_T=\mathbb{H}\otimes\widetilde{\mathbb{H}}$, where the first is the usual Hilbert space while the following is the copy of the first one, in which, obey the commutation (or anti-commutation) relations
\begin{eqnarray}
[\mathcal{A},\mathcal{B}]=if_{ab}^{c}\mathcal{C},\quad\quad[\widetilde{\mathcal{A}},\widetilde{\mathcal{B}}]=-if_{ab}^{c}\widetilde{\mathcal{C}},\quad\quad [\widetilde{\mathcal{A}},\mathcal{B}]=0,
\end{eqnarray}
where $\mathcal{A}$, $\mathcal{B}$ and $\mathcal{C}$ are arbitrary operators in $\mathbb{H}$ and $f_{ab}^c$ are structure constants.

The tilde operators (which are in the $\widetilde{\mathbb{H}}$ space) are related to the non-tilde ones by
\begin{eqnarray}
(\mathcal{A}\mathcal{B})^{\widetilde{}}&=&\widetilde{\mathcal{A}}\widetilde{\mathcal{B}},\quad\quad(\widetilde{\mathcal{A}})^{\widetilde{}}=\varsigma\mathcal{A},\nonumber\\
(a\mathcal{A}+\mathcal{B})^{\widetilde{}}&=&a^{*}\widetilde{\mathcal{A}}+\widetilde{\mathcal{B}},\quad\quad
(\mathcal{A}^\dagger)^{\widetilde{}}=(\widetilde{\mathcal{A}})^\dagger,\label{eq55}
\end{eqnarray}
where $a$ is a complex constant, $\varsigma=1$ for bosons and $\varsigma=-1$ for fermions. Note that the $\mathcal{A}$ operators are the true observables of the theory, while $\widetilde{\mathcal{A}}$ exist only for symmetry purposes. 

In this formalism, there are symmetry generators $\widehat{\mathcal{A}}$ of the thermal space, related to the physical transformations of the system, which are
\begin{eqnarray}
\widehat{\mathcal{A}}=\mathcal{A}-\widetilde{\mathcal{A}},
\end{eqnarray}
in other words, although the tilde and non-tilde operators have the meaning already discussed, it is often necessary to calculate the contribution given by $\widehat{\mathcal{A}}$. Taking the Hamiltonian of the system as an example, it is known that $H$ provides the possible energy states of the system while $\widetilde{H}$ is the tilde-conjugate of the first to duplicate the space and, consequently, composes the thermal space. But only $\hat{H}$ is the operator that gives the time-translational character, i.e., applying it in a certain state gives the temporal evolution of the system.

In addition, by this approach, the average of a dynamical operator $\mathcal{A}$ in thermal equilibrium is given by $\left\langle\mathcal{A}\right\rangle=\bra{0(\beta,\mu)}\mathcal{A}\ket{0(\beta,\mu)}$, where 
\begin{eqnarray}
\ket{0(\beta,\mu)}=\frac{1}{\sqrt{Z(\beta,\mu)}}\sum_n e^{-\beta(E_n-\mu N_n)/2}\ket{n,\widetilde{n}}\label{eq06}
\end{eqnarray}
is the thermal vacuum state, with $Z(\beta,\mu)$ being the grand-canonical partition function, $\mu$ is the chemical potential and $E_n$ and $N_n$ are the energy and occupation number eigenvalues of the $n$-th state, respectively. The temperature $T$ is introduced by the factor $\beta=1/k_BT$, where $k_B$ is the Boltzmann constant.

It should be noted in Eq. (\ref{eq06}) that, as the states $\ket{n}$ and $\ket{\widetilde{n}}$ are in the non-tilde and tilde spaces, respectively, the vector $\ket{n,\widetilde{n}}=\ket{n}\otimes\ket{\widetilde{n}}$ is in the thermal  Hilbert space $\mathbb{H}_T$. Then, to make this link between them, there is another important ingredient of this formalism, the Bogoliubov transformation, which acts by making the connection by uniting thermal and non-thermal operators.

Such transformations permit writing the zero-temperature operators in terms of the $\beta$-dependent ones and vice-versa. Then, looking at the fermion field given by Eq. (\ref{eq07}) these rules can be formulated for its creation and annihilation operators as follows
\begin{eqnarray}
a_s(n,\vec{p}_{\cancel{y}})&=&(1-f)^{1/2}a_s(\beta,n,\vec{p}_{\cancel{y}})+(f)^{1/2}\widetilde{a}^\dagger_s(\beta,n,\vec{p}_{\cancel{y}}), \nonumber\\ a^{\dagger}_s(n,\vec{p}_{\cancel{y}})&=&(1-f)^{1/2}a^{\dagger}_s(\beta,n,\vec{p}_{\cancel{y}})+(f)^{1/2}\widetilde{a}_s(\beta,n,\vec{p}_{\cancel{y}}),\label{eq08}
\end{eqnarray}
for particles, and
\begin{eqnarray}
b_s(n,\vec{p}_{\cancel{y}})&=&(1-f)^{1/2}b_s(\beta,n,\vec{p}_{\cancel{y}})+(f)^{1/2}\widetilde{b}^\dagger_s(\beta,n,\vec{p}_{\cancel{y}}), \nonumber\\ b^{\dagger}_s(n,\vec{p}_{\cancel{y}})&=&(1-f)^{1/2}b^{\dagger}_s(\beta,n,\vec{p}_{\cancel{y}})+(f)^{1/2}\widetilde{b}_s(\beta,n,\vec{p}_{\cancel{y}}),\label{eq09}
\end{eqnarray}
for antiparticles. Here
\begin{eqnarray}
f=f(E_n,\mu)=\frac{1}{e^{\beta(E_n-\mu)}+1},\label{eq15}
\end{eqnarray}
is the Fermi-Dirac distribution. It is important to note that the chemical potential of the particle is different from that of the antiparticle.

In addition to Eqs. (\ref{eq08}) and (\ref{eq09}), we have the anti-commutation relations given by
\begin{eqnarray}
\left\{a_s(\beta,n,\vec{p}_{\cancel{y}}),a^\dagger_{s^\prime}(\beta,n^{\prime},\vec{p^{\prime}}_{\cancel{y}})\right\}=(2\pi)^2\delta_{s,s^{\prime}}\delta_{n,n^{\prime}}\delta(p_x-p^{\prime}_x)\delta(p_z-p^{\prime}_z)\label{eq10}
\end{eqnarray}
for particles and analogously for antiparticles. The others are equal to zero.

Therefore, the interpretation of Eqs. (\ref{eq08}), (\ref{eq09}) and (\ref{eq10}) is very simple, [$b^\dagger_s$]$a^\dagger_s(\beta,n,\vec{p}_{\cancel{y}})$ acts by creating a [anti]fermion with temperature $T$, spin $s$, 3-momentum $\vec{p}=(p_x,p_z)$ and energy $E_n$ at an almost delocalized space-time point (by the uncertainly principle) in which there is a classical magnetic field. In ``almost'', we mean that we do not have information about $x$ and $z$ components but $y$ does, since it is given by spinor dependency and, moreover, even if this operator acts in a situation where $p_y=0$, this component cannot be defined. In this way, the description of all fermionic mechanics in a presence of a background potential at finite temperature becomes straightforward, since $\psi(x^\nu,\beta)$ creates a Dirac particle with temperature $\beta$ at a space-time point $x^\nu$ over a superposition of spin, energy and momentum states. The same concepts can be used for an antiparticle using the $\bar{\psi}(x^\nu,\beta)$ operator.

Now, keeping it in mind, we can extract some important quantities like the differential cross-section for any scattering process.

\subsection{Calculation of the differential cross-section}

In this subsection, the main objective is to calculate the differential cross-section for a scattering process considering two specific cases of the $n$ parameter which is related to the intensity of the external magnetic field. In order to analyze these cases, let us consider the dispersion relation given in Eq. (\ref{eq14}). From this  the following relation is extracted
\begin{eqnarray}
p^\alpha p_\alpha =M^2=E_n^2-|\vec{p}|^2=M^2+p_z^2+2neB-|\vec{p}|^2
\end{eqnarray}
which implies on
\begin{eqnarray}
p_x^2+p_y^2=2neB.\label{eq38}
\end{eqnarray}

Furthermore, taking into consideration Eq. (\ref{eq38}), we can see that this relation forms a circle in the $p_{\perp}$ plane, with a radius of $r=\sqrt{2neB}$. Therefore, when performing integrations over the perpendicular momentum components, which may arise frequently in certain calculations, we find that it is just
\begin{eqnarray}
\int d^2\vec{p^\prime}_\perp=\text{circle area}=\pi r^2=(2\pi)neB.\label{eq44}
\end{eqnarray}
And, for Eq. (\ref{eq38}) two different cases, $n=0$ and $n\neq0$, are considered. In this context, the differential cross-section is investigated.

 \subsubsection{Calculations for $n\neq0$}
 
The differential cross-section is defined as
\begin{eqnarray}
d\sigma=\frac{\text{density of final states}}{\text{flux of incoming particles}}W_{fi},\label{eq37}
\end{eqnarray}
where $W_{fi}$ is the probability density of the incoming state $\ket{i}$ decay into outcome state $\ket{f}$. In a presence of an external field, this quantity is given by 
\begin{eqnarray}
W_{fi}=\frac{(2\pi)^4}{V^4}\delta^4\left(\sum p^{(\text{in})}_{\cancel{y}}-\sum p^{(\text{out})}_{\cancel{y}}\right)\langle|\mathcal{M}_{fi}|^2\rangle,
\end{eqnarray}
since the sums in $p^{(\text{in})}$ and $p^{(\text{out})}$ guarantee the momentum conservation, but with their $y-$components set to zero, and $V $ is a space-time volume stands out for  normalization purposes.

The density of final states, given in Eq. (\ref{eq37}), for a scattering process $p+k\to p^\prime+k^\prime$ in the center of mass frame, where all particle energies are $E_n$, is
\begin{eqnarray}
\text{density}=g_n\frac{V}{(2\pi)^3}\frac{d^3 \vec{p^{\prime}}}{2E_n}g_n\frac{V}{(2\pi)^3}\frac{d^3 \vec{k^{\prime}}}{2E_n}=g_n^2\frac{V^2}{(2\pi)^6}d^2 \vec{p^{\prime}}_{\perp}d^2 \vec{k^{\prime}}_{\perp}\frac{dp_z^{\prime}dk_z^{\prime}}{(2E_n)^2}
\end{eqnarray}
in such a way that the phase space element, in the presence of a magnetic field oriented in the $z$-direction, is given by
\begin{eqnarray}
\frac{d^3\vec{p}}{(2\pi)^3} = \frac{d^2\vec{p}_{\perp}dp_z}{(2\pi)^3}\label{eq45}
\end{eqnarray}
where $\vec{p}_{\perp}$ represents the momentum components $p_x$ and $p_y$, which are orthogonal to the field orientation, and $g_n$ is the spin degeneracy of the energy state, with $g_0=1$ and $g_n=2$ for $n=1,2,3,...$ \cite{dissertmag}.

For what follows, it is usual to write the Mandelstam variables which are given as
\begin{eqnarray}
s=(p+k)^2;\quad\quad t=(p-p^{\prime})^2;\quad\quad u=(p-k^\prime)^2.
\end{eqnarray}
Then, the flux written in Eq. (\ref{eq37}), when the initial particles has mass $M_i$, is
\begin{eqnarray}
\text{flux}=\frac{4}{V^2}\sqrt{\left(p\cdot k\right)^2-M_i^4}=\frac{2}{V^2}\sqrt{s(s-4M_i^2)}.\label{eq41}
\end{eqnarray}
Thus, the differential cross-section Eq. (\ref{eq37}) becomes
\begin{eqnarray}
d\sigma(\beta)=\langle|\mathcal{M}_{fi}(\beta)|^2\rangle g_n^2\frac{\delta^4\left(p_{\cancel{y}}+k_{\cancel{y}}-p^\prime_{\cancel{y}}-k^\prime_{\cancel{y}}\right)}{2\sqrt{s(s-4M_i^2)}}\frac{d^2 \vec{p^{\prime}}d^2 \vec{k^{\prime}}}{(2\pi)^2}\frac{d p^{\prime}_z d k^{\prime}_z}{(2E_n)^2}.\label{eq33}
\end{eqnarray}
The cross-section is obtained from Eq. (\ref{eq33}) by integration and by summing all possibles Landau indices $n$. Performing the integration on $\vec{k}^{\prime}$, the resulting delta function is only on $E$, i.e.,
\begin{eqnarray}
\sigma(\beta)=\sum_n\int \langle|\mathcal{M}_{fi}(\beta)|^2\rangle g_n^2\frac{\delta(E_n^{(p)}+E_n^{(k)}-E_n^{(p^\prime)}-E_n^{(k^\prime)})}{2\sqrt{s(s-4M_i^2)}}\frac{d^2 \vec{p^{\prime}_\perp}d p^{\prime}_z}{(2\pi)^2(2E_n)^2}.\label{eq71}
\end{eqnarray}
However, due to the delta function evaluation, the transition amplitude involves momentum conservation, given by $k^\prime=p+k-p^\prime$. Therefore, Eq. (\ref{eq71}) can be rewritten as 
\begin{eqnarray}
\sigma(\beta)=\sum_n\sigma_n(\beta)\label{eq34}
\end{eqnarray}
with
\begin{eqnarray}
\sigma_n(\beta)=g_n^2\int \langle|\mathcal{M}_{fi}(\beta)|^2\rangle \frac{\delta(E_n^{(p)}+E_n^{(k)}-E_n^{(p^\prime)}-E_n^{(k^\prime)})}{2\sqrt{s(s-4M_i^2)}}\frac{d^2 \vec{p^{\prime}_\perp}d p^{\prime}_z}{(2\pi)^2(2E_n)^2}.\label{eq39}
\end{eqnarray}

Making $E_n^{(p)}+E_n^{(k)}=\sqrt{s}$ and $E_n^{(p^\prime)}+E_n^{(k^\prime)}=2\sqrt{M_f^2+p_z^{\prime 2}+2neB}=E^\prime_n$ in Eq. (\ref{eq39}), and using
\begin{eqnarray}
dE^\prime_n=2\frac{p^\prime_zdp^\prime_z}{\sqrt{M_f^2+p_z^{\prime 2}+2neB}}=\frac{E^{\prime}_n}{E_n^2}p^\prime_zdp^\prime_z,
\end{eqnarray}
the $n$-th cross-section is written as
\begin{eqnarray}
\sigma_n(\beta)=\frac{g_n^2}{(4\pi)^2}\int \frac{\langle|\mathcal{M}_{fi}(\beta)|^2\rangle}{2\sqrt{s(s-4M_i^2)}}\delta(\sqrt{s}-E^\prime_n)\frac{d^2 \vec{p^{\prime}_\perp}dE^\prime_n}{p^\prime_zE^\prime_n}=\frac{g_n^2}{(4\pi)^2}\int\frac{\langle|\mathcal{M}_{fi}(\beta)|^2\rangle}{2sp^\prime_z\sqrt{s-(2M_i)^2}}d^2 \vec{p^{\prime}_\perp},\label{eq36}
\end{eqnarray}
where $M_f$ is the mass of the final fermion. As the field strength reaches very high values, $\mathcal{M}_{fi}$ becomes increasingly independent of the perpendicular momenta, allowing the direct application of the integration in Eq. (\ref{eq44}) to Eq. (\ref{eq36}).

As already discussed, for the treated problem to be physical it is necessary that $E^2-M^2\geq 2neB$. In other words, a limit for $n$ can be set, i.e.,
\begin{eqnarray}
n_{\text{max}}=\text{int}\left\{\frac{E^2-M^2}{2eB}\right\},\label{eq69}
\end{eqnarray}
in such a manner that in the limit of the strongest possible magnetic field, we get $n_{\text{max}}=0$. Then, the total cross-section Eq. (\ref{eq34}) is written as
\begin{eqnarray}
\sigma(\beta)=\sum_{n=0}^{n_{\text{max}}}\sigma_n(\beta).\label{eq40}
\end{eqnarray}

In the limit without field, $B\to0$, the spacing between each $E_n$ and $E_{n+1}$ becomes smaller and smaller. Therefore, the energy levels become a continuum spectrum and $g_n\equiv2$ for all $n$. Then, Eq. (\ref{eq40}) reads
\begin{equation}
\sigma_{B\to0}(\beta)=\int_{0}^{n_{\text{max}}}\sigma_n(\beta)dn,
\end{equation}
and using Eq. (\ref{eq36}), the total cross-section will be 
{\begin{eqnarray}
\sigma_{B\to0}(\beta)&=&\lim_{B\to0}\frac{g_n^2}{(4\pi)^2}\frac{1}{2s\sqrt{s-(2M_i)^2}}\int\left[\int_0^{n_{\text{max}}}\langle|\mathcal{M}_{fi}(\beta)|^2\rangle\frac{n}{\sqrt{E_n^2-M_f^2-2neB}}dn\right]d^2 \vec{p^{\prime}_\perp}\label{eq61}
\end{eqnarray}
which is strongly dependent on the form of $\langle|\mathcal{M}_{fi}|^2\rangle$, i.e., the functional dependence of $B$ and $n$. The last dependence is given uniquely by the form of spin completeness relations Eq. (\ref{eq27}) and Eq. (\ref{eq28}).

\subsubsection{Calculations for $n=0$}
 
When $n=0$ it is necessary to look for a special case. From Eq. (\ref{eq38}) it is extracted $p_x=p_y=0$. In other words, if the situation $n_{\text{max}}=0$ is taken, the momentum of the resulting particles can only be in $p_z$ one-dimensional real space. Therefore, the density of the final states coming from the phase space element is
\begin{eqnarray}
 \text{density}=g_0\frac{V}{(2\pi)}\frac{dp^{\prime}_z}{2E_0}g_0\frac{V}{(2\pi)}\frac{dk^{\prime}_z}{2E_0}=\frac{V^2}{(2\pi)^2}\frac{dp^{\prime}_zdk^{\prime}_z}{(2E_0)^2},
\end{eqnarray}
and the probability density of the scattering process becomes
\begin{eqnarray}
W_{fi}=\frac{(2\pi)^2}{V^4}\delta\left(\sum p_z^\text{(in)}-\sum p_z^\text{(out)}\right)\delta\left(E_n^{(p)}+E_n^{(k)}-E_n^{(p^\prime)}-E_n^{(k^\prime)}\right)\langle|\mathcal{M}_{fi}|^2\rangle.
\end{eqnarray}
The flux of incoming particles is the same as given in Eq. (\ref{eq41}). Therefore, in this present case, we get
\begin{eqnarray}
\sigma_0(\beta)&=&\int\frac{\langle|\mathcal{M}_{fi}(\beta)|^2\rangle}{2\sqrt{s(s-4m_e^2)}}\delta\left(E_n^{(p)}+E_n^{(k)}-E_n^{(p^\prime)}-E_n^{(k^\prime)}\right)\frac{dp^\prime_z}{(2E_0)^2}\nonumber\\
&=&\int\frac{\langle|\mathcal{M}_{fi}(\beta)|^2\rangle}{8\sqrt{s(s-4m_e^2)}}\delta(\sqrt{s}-E^\prime_n)\frac{dE^\prime_n}{p^\prime_zE^\prime_n}=\frac{\langle|\mathcal{M}_{fi}(\beta)|^2\rangle}{8p^\prime_z s\sqrt{s-4m_e^2}}.\label{eq42}
\end{eqnarray}

In short, any $1+2\to3+4$ scattering process carried out at finite temperature in the center of mass frame subject to an external magnetic field will have the total cross-section given by Eq. (\ref{eq40}), where $\sigma_n(\beta)$ is given by Eq. (\ref{eq36}) for $n>1$ and $\sigma_0(\beta)$ is given in Eq. (\ref{eq42}). When the field strength $B$ becomes negligible, $n_{\text{max}}\to\infty$, then the energy levels become a continuum and $\sigma_n$ needs to be integrated.

On the other hand, when the scattering process is analyzed over a very strong field the only allowed level, with respect to Eq. (\ref{eq69}), is the one where $n_{\text{max}}=0$. Then the only quantity needed to be taken into account is $\sigma_0(\beta)$.

In the next section, the meaning of the chemical potential and its importance for the scattering process are investigated.

\section{The Chemical Potential}\label{secpotential}

Starting from the TFD formalism, one can observe the dependence of a physical system on temperature and chemical potential, as can be seen in Eq. (\ref{eq06}). Note that temperature is a well-known quantity, whereas chemical potential does not have a well-understood physical picture either in basic physics or in more complex studies of quantum field theory \cite{understanding}. In this way, first we need to explain the meaning of this quantity and, later, what it represents in QED processes.

\subsection{The meaning of the chemical potential}

In  Eqs. (\ref{eq08}) and (\ref{eq09}) the chemical potential $\mu$ appears due to Eq. (\ref{eq15}). Although almost never discussed, the role of chemical potential in QED processes is very important \cite{comptonfinite}. This quantity comes from the free energy related to particle changes. Since the vacuum is not empty, such as real particles, there are a very large number of liquid particles present in any real scattering process that are created and subsequently annihilated without violating the uncertainly principle \cite{iengoquantum}.

The chemical potential in a system of fermionic particles is equal to the Fermi energy at zero temperature. Where is the highest energy level of the system at the fundamental state \cite{landaustat}. In addition, the study of the contribution of this term is very important in several cases where it is necessary to consider the macroscopic behavior of the problem, i.e., the thermodynamic connection.

As the grand-canonical formalism of statistical physics is used, it is known that the connection with thermodynamics is made by the link $Z(\beta,\mu)\to\Phi(T,\mu)$, where $\Phi$ is the thermodynamic grand-potential \cite{callen}. In this way, it is known that the chemical potential is the free energy related to the changes of the particles. As in the present discussion the Dirac field is treated, there are two types of particles, with positive and negative energy. Then, can be defined
\begin{eqnarray}
N_{+}=\frac{\partial\Phi}{\partial \mu_{+}},\quad\quad\quad N_{-}=\frac{\partial\Phi}{\partial \mu_{-}}.
\end{eqnarray}
It is important to note that the annihilation of particles implies the creation of antiparticles and vice-versa, such that $dN_{+}=-dN_{-}$. Therefore defining $\mu$ as the chemical potential of the system we get $\mu_{+}\equiv\mu$ and $\mu_{-}=-\mu$, for fermions and anti-fermions, respectively.

\begin{figure}[ht]
\subfigure[aa][The distribution function of a particle at close-to-zero temperature (black solid line) and $T>0$ (red dashed line) cases.]{\includegraphics[width=8.1cm]{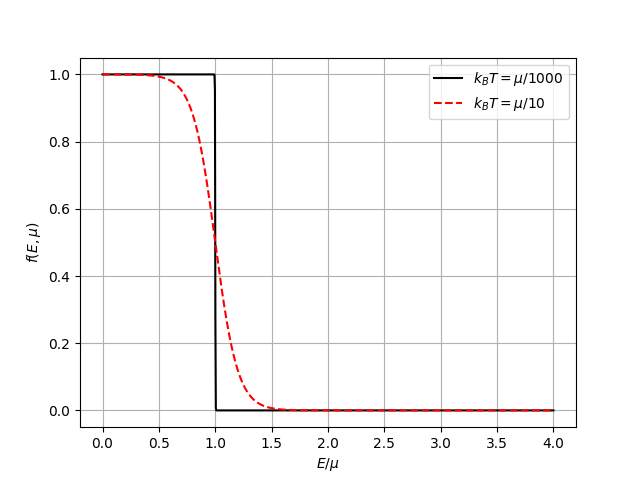}\label{fig2a}}
\subfigure[bb][The distribution function of an antiparticle at close-to-zero temperature (black solid line) and $T>0$ (red dashed line) cases.]{\includegraphics[width=8.1cm]{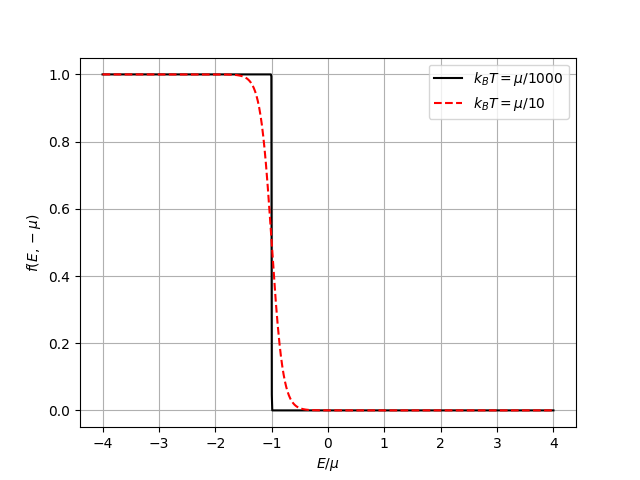}\label{fig2b}}
\subfigure[cc][The distribution function of an antiparticle with the change $-E\to\epsilon$, at close-to-zero temperature (black solid line) and $T>0$ (red dashed line) cases.]{\includegraphics[width=8.1cm]{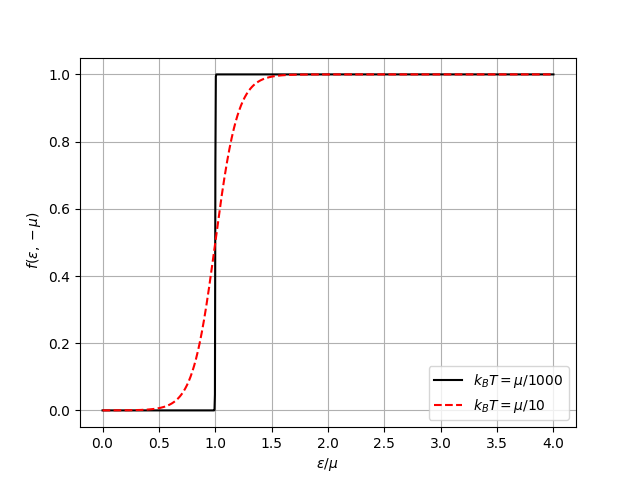}\label{fig2c}}
\caption{Distribution functions in terms of the Energy.}\label{fig2}
\end{figure}

The distribution function $f(E,\mu)$ given in Eq. (\ref{eq15}) is plotted in Figure \ref{fig2}. The Figure \ref{fig2a} is the well-known Fermi-Dirac function that describes the population of energy states of a given system with chemical potential $\mu$ in two different temperature cases. At zero temperature, all states with energy less than $\mu$ are filled and empty otherwise. When the temperature increases, the forward states become filled. Looking at Figure \ref{fig2b}, we notice that for antiparticles, although the step is moved to $-\mu$, the profile is similar to the previous graph and the same analysis can be applied to this case. The last function shown in Figure \ref{fig2c} is the same as the previous one, but with the redefinition of the energy by $\epsilon=-E$, and this approach displays an electron distribution like the first figure.

In a quantum field theory, such as QED, it is possible to associate a chemical potential with each conserved quantum number such as the lepton \cite{dominik}. For our present purposes, the lepton number density is written as
\begin{eqnarray}
\rho =N_{+}-N_{-}= \sum_{k}g_k\int\frac{d^3\vec{p}}{(2\pi)^3}\left[f(E_k(p_z),\mu)-f(E_k(p_z),-\mu)\right],\label{eq16}
\end{eqnarray} 
where the Fermi-Dirac distribution given in Eq. (\ref{eq15}) is used \cite{comptonfinite, landaustat}. The $k$-sum and the $\vec{p}$-integration ensure that all energy eigenstates, Eq. (\ref{eq14}), need to be taken into account. Then,  Eq. (\ref{eq16}) can be written as
\begin{eqnarray}
\rho=\sum_n g_n\int\frac{d^3\vec{p}}{(2\pi)^3}\left[\frac{1}{\zeta^{-1} e^{\beta\sqrt{M^2+p_z^2+2neB}}+1}-\frac{1}{\zeta e^{\beta\sqrt{M^2+p_z^2+2neB}}+1}\right],\label{eq17}
\end{eqnarray}
with $\zeta=e^{\beta\mu}$ being the fugacity of the system. At this moment, it is again necessary to make a distinction between the cases $n\neq0$ and $n=0$. For the former, looking at Eq. (\ref{eq45}) and Eq.  (\ref{eq44}), we get from Eq. (\ref{eq17})
\begin{eqnarray}
\rho=\sum_n(2neB)g_n\int_{-\infty}^{\infty}\frac{dp_z}{(2\pi)^2}\left[\frac{1}{\zeta^{-1} e^{\beta\sqrt{M^2+p_z^2+2neB}}+1}-\frac{1}{\zeta e^{\beta\sqrt{M^2+p_z^2+2neB}}+1}\right]\equiv\sum_n\rho_n.\label{eq46}
\end{eqnarray}
Then, it is necessary to calculate all $n$-states densities and add them together.

For the case $n=0$,  Eq. (\ref{eq17}) is written as
\begin{eqnarray}
\rho=\int_{-\infty}^{\infty}\frac{dp_z}{2\pi}\left[\frac{1}{\zeta^{-1} e^{\beta\sqrt{M^2+p_z^2}}+1}-\frac{1}{\zeta e^{\beta\sqrt{M^2+p_z^2}}+1}\right].\label{eq47}
\end{eqnarray}
Due to the constraint imposed by the intense magnetic field on the particle's motion in the $z$-direction, the integration in Eq. (\ref{eq47}) is one-dimensional, unlike Eq. (\ref{eq46}). In addition, both cases of lepton densities have very different forms, and this is due to the strong dependence of the state density on the Landau levels, which as already discussed are finite numbers determined by the amplitude of the external magnetic field. Furthermore, it is interesting to reinforce that, in the case of very strong magnetic fields, only the $n=0$ level is allowed. Consequently, in this regime, the energy does not depend on the magnetic field $B$, and, as a result, neither does the fermion density $\rho$. The chemical potential $\mu$ is determined by inverting the relation between $\rho$ and the parameters $\beta$ and $\zeta$ as given in Eqs. (\ref{eq46}) and (\ref{eq47}) resulting in $\mu=\mu(\beta,\rho)$. Consequently, when considering $n=0$, $\mu$ is independent of the magnetic field. However, for $n\neq0$, $\mu$ becomes dependent on the magnetic field. In other words, in the regime where the energy does not depend on the external field, the chemical potential behaves similarly. Conceptually, we can view $\mu$ as the energy required to add a particle to the fermionic system at zero temperature. Although Eqs. (\ref{eq46}) and (\ref{eq47}) cannot be solved analytically due to the form of the dispersion relation, they can be used to determine the chemical potential as a function of temperature and lepton density.  But in the following, we only look at the latter case. Furthermore, it is interesting to do some analysis on Eq. (\ref{eq47}) looking at the limiting cases, i.e., extracting information when the system is subject to high and low temperatures. For high temperatures (or small values of $\beta$),  Eq. (\ref{eq47}) will be strongly dependent on the value of $\mu$. When the charges involved are symmetrically related, we get $\rho=0$ and therefore $\mu=0$. 

In the low-temperature limit ($\beta\to\infty$), the first term in square brackets in Eq. (\ref{eq47}) is dominant upon its subsequent one because of the exponential present in the denominator (this fact can be seen in Figures \ref{fig2a} and \ref{fig2b}). Then, one can write
\begin{eqnarray}
\rho=\frac{1}{\pi}\int_{0}^{\infty}\frac{dp_z}{e^{\beta (E-\mu)}+1}\label{eq48}.
\end{eqnarray}
It is useful to write it in terms of energy, i.e., putting $dp_z$ in terms of $dE$ and shifting the displacement $E\to E+M$. Thus
\begin{eqnarray}
\rho=\frac{1}{\pi}\int_{0}^{\infty}dE\frac{E+M}{\sqrt{E^2+2EM}}\frac{1}{e^{\beta (E+M-\mu)}+1}\equiv\int_{0}^{\infty}G(E)f(E,\mu-M)dE,\label{eq18}
\end{eqnarray}
with $G(x)=(x+M)/(\pi\sqrt{x^2+2Mx})$. Integrating Eq. (\ref{eq18}) by parts, we get
\begin{eqnarray}
\int_{0}^{\infty}G(E)f(E,\mu-M)dE=-\int_{0}^{\infty}\left[\int_0^{E}G(\varepsilon)d\varepsilon\right]\left(\frac{\partial f}{\partial E}\right)dE,
\end{eqnarray}
and expanding the term in the integrand around $E=\mu-M$, we have
\begin{eqnarray}
\int_0^{E}G(\varepsilon)d\varepsilon=\int_0^{\mu-M}G(\varepsilon)d\varepsilon+\sum_{k=1}^{\infty}\frac{(E-\mu+M)^k}{k!}\frac{d^k}{dE^k}\left[\int_0^{E}G(\varepsilon)d\varepsilon\right]_{E=\mu-M}.
\end{eqnarray}

In this temperature regime, the Sommerfeld expansion can be used,
\begin{eqnarray}
\int_0^{\infty}G(E)f(E,\mu-M)dE=\int_0^{\mu-M}G(\varepsilon)d\varepsilon+\frac{\pi^2}{6}\left(\frac{1}{\beta}\right)^{2}G^\prime(\mu-M)+\mathcal{O}\left[\left(\frac{1}{\beta}\right)^4\right],\label{eq19}
\end{eqnarray}
where $G^{\prime}(x)$ is the derivative of $G(\varepsilon)$ applied to $\varepsilon=x$ \cite{ashcroft}. Then, Eq. (\ref{eq18}) up to second order becomes
\begin{eqnarray}
\rho=\frac{1}{\pi}\left[\sqrt{\mu^2-M^2}-\frac{\pi^2}{6}\left(\frac{1}{\beta}\right)^{2} \frac{M^2}{\left(\mu^2-M^2\right)^{3/2}}\right].\label{eq20}
\end{eqnarray}
Note that Eq. (\ref{eq20}) is a way to determine the chemical potential of this system in terms of the $\rho_N$ and $T$ over a sum of all possible energy states $E_n$. At zero temperature, we get 
\begin{eqnarray}
\mu_0=M\cosh{\theta_0},\quad\quad\text{where}\quad\quad
\rho_0=\frac{M}{\pi}\sinh{\theta_0},\label{eq43}
\end{eqnarray}
similarly to that obtained by \cite{comptonfinite}, but in the context of the presence of a background field. Furthermore, to obtain an expression for the temperature-dependent chemical potential it is necessary to take the following thermodynamical relation \cite{marder}
\begin{eqnarray}
\frac{\partial \mu}{\partial T}=\left.-\left({\frac{\partial \rho}{\partial T}}\right/{\frac{\partial \rho}{\partial\mu}}\right),
\end{eqnarray}
which implies (up to second order on temperature) in
\begin{eqnarray}
\frac{\partial\mu}{\partial T}=\frac{k_B^2M^2\pi^2}{3}\frac{T}{\mu(\mu^2-M^2)}.
\end{eqnarray}

Then, the temperature-dependent chemical potential, up to second order, is given by
\begin{eqnarray}
\mu(T)=\left[M^2+\sqrt{(\mu_0^2-M^2)^2+\frac{2}{3}\left(k_BM\pi\right)^2T^2}\right]^{1/2}\label{eq53}
\end{eqnarray}
or, expanding once more in the Taylor series,
\begin{eqnarray}
\mu(T)=\mu_0+\frac{k_B^2M^2\pi^2}{6\mu_0(\mu_0^2-M^2)}T^2.
\end{eqnarray}

On the other hand, evidencing the Boltzmann factor $e^{-\beta E}$, Eq. (\ref{eq47}) becomes
\begin{eqnarray}
\rho=\int_{0}^{\infty}\frac{dp_z}{\pi}e^{-\beta\sqrt{M^2+p_z^2}}\left[\frac{\zeta}{\zeta e^{-\beta\sqrt{M^2+p_z^2}}+1}-\frac{\zeta^{-1}}{\zeta^{-1} e^{-\beta\sqrt{M^2+p_z^2}}+1}\right]
\end{eqnarray}
and, since $e^{-x}<1$ for $x>0$ the term in square brackets can be expanded into power series, then 
\begin{eqnarray}
\rho=\sum_{k=1}^{\infty}(-1)^{k+1}\int_0^{\infty}\frac{dp_z}{\pi}e^{-\beta k\sqrt{M^2+p_z^2}}\left(\zeta^k-\zeta^{-k}\right).\label{eq49}
\end{eqnarray}
As the integrand of Eq. (\ref{eq49}) is exponentially damped, it is only necessary to take a few first values of $k$. Then, doing just $k=1$, we get
\begin{eqnarray}
\rho=\frac{2}{\pi}\sinh{\beta\mu}\int_0^{\infty}dp_ze^{-\beta\sqrt{M^2+p_z^2}},
\end{eqnarray}
which, through an expansion in temperature, can be summarized in two different regimes,
\begin{eqnarray}
\rho=\frac{2}{\pi\beta}\sinh{\beta\mu} \quad\quad (\beta M<<1)\label{eq50}
\end{eqnarray}
and
\begin{eqnarray}
\rho=\sqrt{\frac{2M}{\pi\beta}}e^{-\beta M}\sinh{\beta\mu} \quad\quad (\beta M>>1).\label{eq51}
\end{eqnarray}
These results are similar to those obtained in \cite{early}. The comparison between $\beta M$ and the unity shows us how greater the factor $k_BT$ is in terms of the rest energy $mc^2$.

Expanding Eq. (\ref{eq50}) and Eq. (\ref{eq51}) we get
\begin{eqnarray}
\mu=\frac{\pi}{2}\rho\quad\quad (\beta M<<1)
\end{eqnarray}
and
\begin{eqnarray}
\mu(T)=M-\frac{1}{2}k_BT\ln{\left(\frac{k_BT}{M}\right)}+\frac{\alpha}{2}k_BT\quad\quad (\beta M>>1)\label{eq52}
\end{eqnarray}
where $\alpha$ is a constant, which depends on the leptonic density of the system.

At very high temperatures, $\beta\to0$ or $\beta M<<1$, as discussed $\rho=0$, therefore $\mu=0$. The result obtained in Eq. (\ref{eq52}) is very interesting and is similar to the one obtained without external field by \cite{comptonfinite}. In addition, as Eq. (\ref{eq52}) can be taken in the low temperatures regime, i.e. $\beta M>>1$, we get a behavior similar to those obtained in Eq. (\ref{eq53}) obtained in Sommerfeld formalism, but making $\mu_0=M$ and taking very low values of $T$. When the temperature increases, both expressions become very different and this is because to obtain Eq. (\ref{eq53}) we assume many approximations that do not happen in the calculation of Eq. (\ref{eq52}).

\begin{figure}[ht]
\includegraphics[scale=0.7]{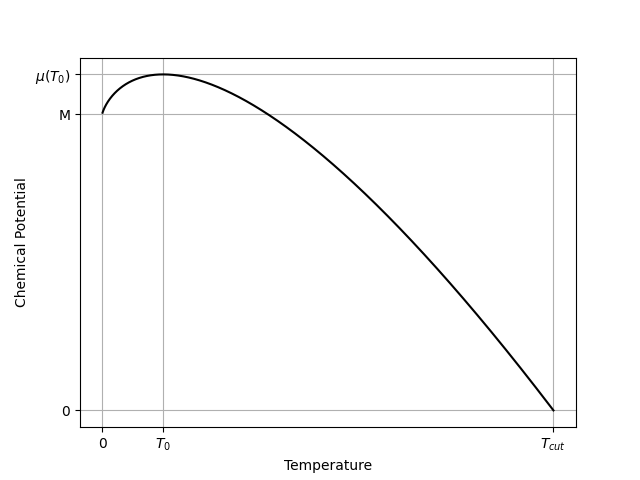}
\caption{Chemical potential function described by Eq. (\ref{eq62}).}
\label{fig5}
\end{figure}

From this, the functional form of the lepton density chemical potential can be summarized as
\begin{eqnarray}
\mu(T)=\begin{cases} 
M-\frac{1}{2}k_BT\ln{\frac{k_BT}{M}}+\frac{\alpha}{2}k_BT, \quad &\text{if} \quad T<T_{\text{cut}}\\
0,\quad &\text{if}\quad T\geq T_{\text{cut}}
\end{cases}\label{eq62}
\end{eqnarray}
whose maximum value occurs for 
\begin{eqnarray}
T_0=\frac{M}{k_B}e^{\alpha-1}.\label{eq70}
\end{eqnarray}
This function is plotted in Figure \ref{fig5}.

In the next subsection, an interpretation of the chemical potential in a QED scattering process is presented.

\subsection{The interpretation of the chemical potential in a QED scattering reaction}

In ordinary thermodynamics, the chemical potential is defined as
\begin{eqnarray}
\mu=\left(\frac{\partial U}{\partial N}\right)_{S,V}\approx U(N+1)-U(N),\label{eq54}
\end{eqnarray}
where $U$ is the internal energy of the system. This quantity is the energy needed to add a particle to the system keeping entropy and volume constant \cite{callen}. 

In a QED system surrounded by thermal and particle reservoirs, the particle population can be described by the functions given in Figures \ref{fig2a} and \ref{fig2c}, i.e., there are two types of fermions, with positive and negative energy, as shown in Figure \ref{fig3a}. The existence of particles with negative energy can be understood with the Fermi sea formalism \cite{dirac}. In this way, the antiparticle is understood as if there were ``holes'' in the distribution of negative-energy fermions. The processes of annihilation and pair creation can be explained by the hole-fermion interaction. Pair creation occurs when a negative energy fermion absorbs an external photon, it jumps to an empty state leaving an empty space where it was, that is, a hole with positive energy and charge, a anti-fermion. On the other hand, the pair annihilation process happens when the positive energy fermion decays into an empty space in the negative energy spectrum emitting a photon with energy equal to the sum of the energies of the fermion and the hole.

\begin{figure}[ht]
\subfigure[aaaa][The particle population of the system where the energy range of the fermions with the highest positive and negative energies is $\Delta E=2\mu_0$.]{\includegraphics[width=8.6cm]{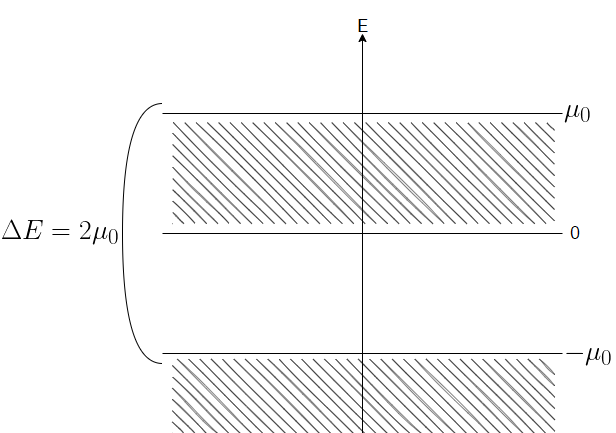}\label{fig3a}}
\subfigure[sssss][The particle population of the system when the highest negative fermion absorbs an external photon with energy $E_\gamma>2\mu_0$.]{\includegraphics[width=7.1cm]{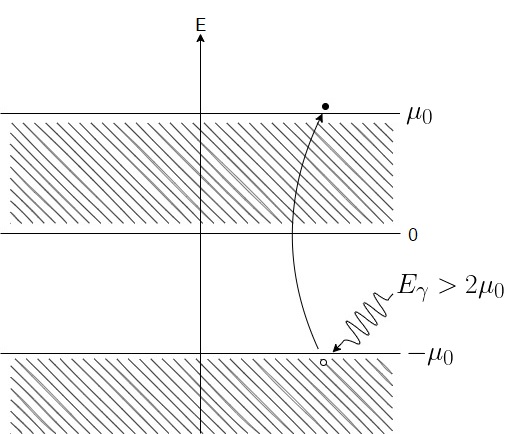}\label{fig3b}}
\caption{Schemes for energy spectra of electrons at $T=0$ temperature.}\label{fig3}
\end{figure}

Figure \ref{fig3a} shows the energy spectra of the system when the distribution is given by \ref{fig2a} and \ref{fig2b}. In this scheme, we can see that in the energy range $(0,\mu_0)$ such as $(-\infty,-\mu_0)$ are fully filled while the interval $(-\mu_0,0)$ is a forbidden zone. Furthermore, the interpretation of Eq. (\ref{eq54}) shows us that at zero temperature $\mu_0$ is the energy required to add a fermion to the system. It is important to notice that, to remove the higher energy particle it is necessary to remove a fermion with energy $\mu_0$ or, in other words, add an $E=-\mu_0$ to the system. Therefore, the latter implies the addition of a negative energy fermion to the system.

In terms of the pair creation and annihilation processes, since the antiparticle is a hole in the negative energy fermion distribution, it can only exist if the excited negative energy particle jumps to a state where $E>\mu_0$, where are only the free states. In this point of view, it is necessary for the fermion to absorb an external photon with energy greater than the separation interval between the states, as shown in Figure \ref{fig3b}. In short, anti-fermions cannot be created if the received energy is not greater than $2\mu_0$.

The idea behind the use of the grand-canonical formalism is that the vacuum can behave like a reservoir of particles. This approach lies in the fact that infinite virtual processes happen in a vacuum, that is, quanta of particles are created and annihilated all the time without violating the energy-time uncertainly principle \cite{iengoquantum}. But when a virtual pair absorbs an external photon, it becomes real, so this fact configures the vacuum as a reservoir.

Figure \ref{fig4} shows a pictorial scheme of the QED system connected with both reservoirs. Special attention is required to the particle reservoir. Note that it is impossible to conceive an antiparticle, with energy $E<\mu_0$, in a physical reaction, since there are no negative energy fermions in the forbidden zone. In addition, it is not possible to create another particle where $E<\mu_0$, since all states are filled. Therefore, for any particle-antiparticle reaction, it is necessary that the condition $E>\mu_0$ is satisfied.

\begin{figure}[ht]
\includegraphics[scale=0.35]{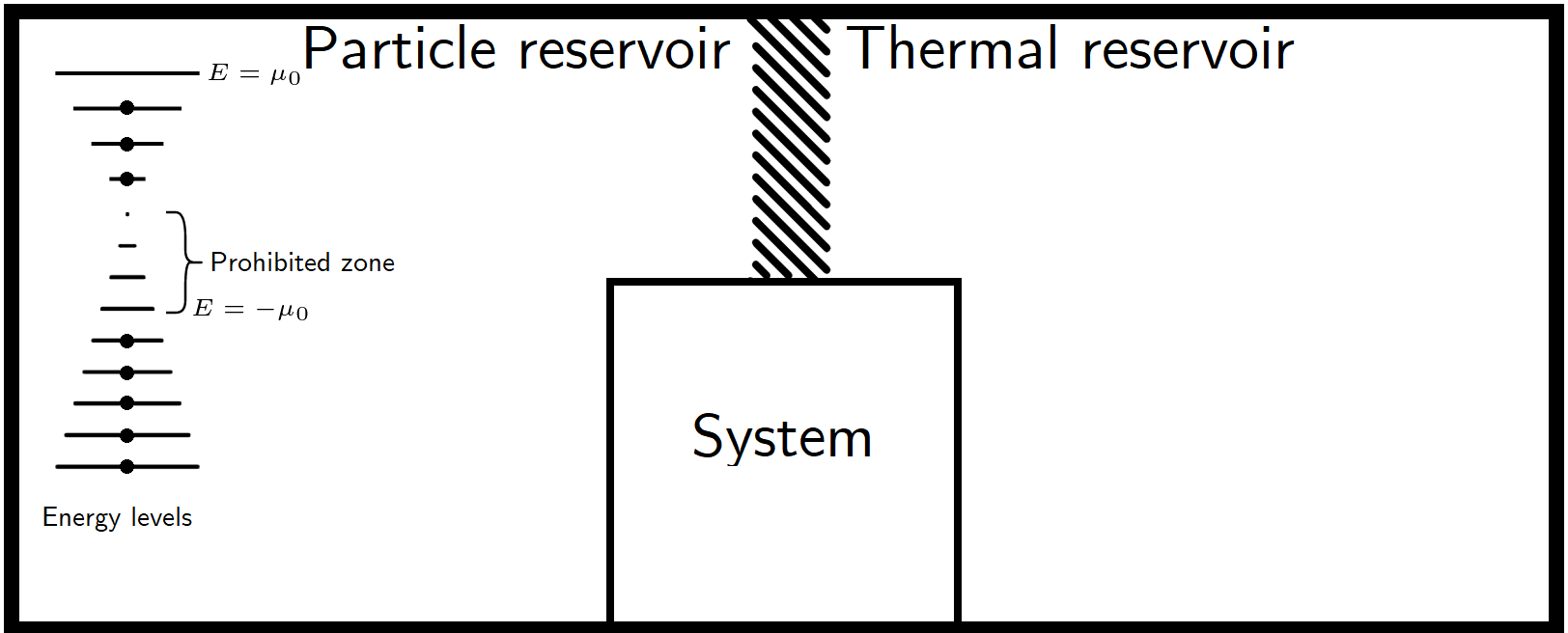}
\caption{Scheme of a connected system with two reservoirs, one for particles and the other for a thermal bath. In the former, the energy spectrum of the particles is represented by the energy levels placed on the left.}\label{fig4}
\end{figure}

Furthermore, following the results given in Eq. (\ref{eq43}) or Eq. (\ref{eq52}), the chemical potential of a QED system at zero temperature is just the mass of the fermion that composes it. This fact can be easily verified in a practical manner. It would be reasonable to think that the minimum energy needed to create a particle is that which keeps the particle at rest.

On the other hand, in a system composed of fermions of different flavors, we get from Eq. (\ref{eq54}),
\begin{eqnarray}
\mu_i=\left(\frac{\partial U}{\partial N_i}\right)_{S,V,N_j}\approx U(N_i+1)-U(N_i),
\end{eqnarray}
where the indices $i$ and $j$ ranges cover all flavors with $i\neq j$. Then, looking at the $e^{+}e^{-}\to l^{+}l^{-}$ reaction, it is necessary to consider the electron and lepton distribution functions. That is, there are two reservoirs, one for each particle.

For bosons, the discussion is a little different, because at zero temperature the energy necessary to add a bosonic particle to the system is that in which  the Bose-Einstein condensate are. Especially for photons, which are massless and non-interacting particles, we can return to the black-body radiation problem, where we have $\mu=0$ for all temperatures \cite{landaustat}.

Finally, we can say that a QED reaction, at zero temperature, can only occur if the total energy of the initial state is greater than $2\mu_0$. When the temperature assumes very high values, i.e., greater than $T_0$, given by Eq. (\ref{eq70}), the interpretation of the chemical potential no longer makes sense, since that $\mu$ can be defined as zero. Therefore, all important quantities in the limit $\beta\to0$ do not depend on a comparison between $E$ and $\mu$.

Up to now, we have seen how to describe a fermion under the effect of a classical background magnetic field at finite temperature, what is chemical potential, and what it represents in a reaction. In the next section, the QED reactions are revised and it is shown which features become relevant when we take these phenomena into account.

\section{The scattering process}\label{secscattering}

The QED Lagrangian for an arbitrary lepton with mass $m_l$ is given by
\begin{eqnarray}
\mathcal{L}=\bar{\psi}_l\left(iD_\nu\gamma^\nu-m_l\right)\psi_l-\frac{1}{4}F_{\alpha\nu}F^{\alpha\nu},\label{eq01}
\end{eqnarray}
with $D_\nu=\partial_\nu-ie A_\nu$ being the covariant gauge derivative. It is important to note that, in the presence of a background field, the potential is written as
\begin{eqnarray}
A_\nu(x)=A_{\nu}^{(D)}(x)+A_{\nu}^{(B)}(x),\label{eq02}
\end{eqnarray}
where $A_\nu^{(D)}$ and $A_\nu^{(B)}$ are the dynamical and background fields, respectively. In other words, while  the first quantity is the free field of a propagating photon, the second is the potential vector of classical background magnetic field, as in Eq. (\ref{eq21}).

Note from Eq. (\ref{eq02}) that $F_{\alpha\nu}=F_{\alpha\nu}^{(D)}+F_{\alpha\nu}^{(B)}$, with
\begin{eqnarray}
F_{\alpha\nu}^{(D)}=\partial_\alpha A_\nu^{(D)}-\partial_\nu A_\alpha^{(D)} \quad\quad\mathrm{and}\quad\quad\quad F_{\alpha\nu}^{(B)}=\partial_\alpha A_\nu^{(B)}-\partial_\nu A_\alpha^{(B)},
\end{eqnarray}
where due to the structure of the potential field, $F_{\alpha\nu}^{(B)}$ is constant. Then the contribution of this term is a constant of motion and does not affect the field equations \cite{nivaldo}. Therefore, the Lagrangian given in Eq. (\ref{eq01}) becomes
\begin{eqnarray}
\mathcal{L}&=&\bar{\psi}_l\left[i(\partial_\nu-ieA_\nu^{(B)})\gamma^\nu-m_l\right]\psi_l+e\bar{\psi}_lA_\nu^{(D)}\gamma^\nu\psi_l-\frac{1}{4}F_{\alpha\nu}^{(D)}F^{\alpha\nu}_{(D)}\nonumber\\
&=&\bar{\psi}_l\left(iD_\nu\gamma^\nu-m_l\right)\psi_l-\frac{1}{4}F_{\alpha\nu}^{(D)}F^{\alpha\nu}_{(D)},\label{eq03}
\end{eqnarray}
where $D_\nu=-i\Pi_\nu-ieA_\nu^{(D)}$ and $\Pi_\nu$ is the same as given by Eq. (\ref{eq04}).

Therefore, from Eq. (\ref{eq04}), Eq. (\ref{eq03}) can be understood as the theory describing the usual QED, but the dynamics of a Dirac field is now formulated for a lepton with electric charge $-e$ and mass $m_l$ in the presence of a constant magnetic field. Thus, the interaction with the gauge photon can be described by a minimal coupling. In other words, the Feynman rules can be applied to the present case.

From Eq. (\ref{eq03}) the interaction Lagrangian is given as
\begin{eqnarray}
\mathcal{L}_{\text{int}}(x)=e\bar{\psi}_l(x)\gamma^\nu\psi_l(x) A_\nu^{(D)}(x).
\end{eqnarray}
Using this Lagrangian the term of second order of the scattering matrix is written as
\begin{eqnarray}
\widehat{S}^{(2)}&=&\frac{(-i)^2}{2}\int d^4x_1d^4x_2:\widehat{\mathcal{L}}(x_1)\widehat{\mathcal{L}}(x_2):,
\nonumber\\&=&\frac{(-i)^2}{2}\int d^4x_1d^4x_2:\left[\mathcal{L}(x_1)-\widetilde{\mathcal{L}}(x_1)\right]\left[\mathcal{L}(x_2)-\widetilde{\mathcal{L}}(x_2)\right]:, \label{eq23}
\end{eqnarray}
where $:(\quad):$ is the ordering operator. It is important to note that only non-tilde operators are related to measurable physical quantities, therefore, after performing the Bogoliubov transformations to make this theory thermal, one can cut the tilde creation and annihilation operators in Eq. (\ref{eq23}).

For the scattering process $e^{+}e^{-}\to l^{+}l^{-}$ in the presence of a background magnetic field within the TFD formalism, the initial and final states are described as 
\begin{eqnarray}
\ket{i(\beta,\mu)}=\sqrt{2E_n}\sqrt{2E_j}a^\dagger_s(\beta,n,\vec{p}_{\cancel{y}})b^\dagger_\lambda(\beta,j,\vec{k}_{\cancel{y}})\ket{0(\beta,\mu)},\nonumber\\
\ket{f(\beta,\mu)}=\sqrt{2E_{n^\prime}}\sqrt{2E_{j^\prime}}a^\dagger_{s^\prime}(\beta,n^{\prime},\vec{p^\prime}_{\cancel{y}})b^\dagger_{\lambda^\prime}(\beta,j^{\prime},\vec{k^{\prime}}_{\cancel{y}})\ket{0(\beta,\mu)},\label{eq22}
\end{eqnarray}
where the pre-factors $\sqrt{2E}$ are present for normalization purposes \cite{peskin}.

Therefore, the probability amplitude of this scattering process to occur is given by
\begin{eqnarray}
\mathcal{M}_{fi}&=&\bra{f(\beta,\mu)}\widehat{S}^{(2)}\ket{i(\beta,\mu)}\nonumber\\
&=&-\frac{e^2}{2}\int d^4x_1d^4x_2\bra{f}:\left\{\left[\bar{\psi}_e(x_1)\gamma^\alpha\psi_e(x_1)A_\alpha^{(D)}(x_1)\right]^{\widetilde{}}-\bar{\psi}_e(x_1)\gamma^\alpha\psi_e(x_1)A_\alpha^{(D)}(x_1)\right\}\nonumber\\&\times &\left\{\left[\bar{\psi}_l(x_2)\gamma^\nu\psi_l(x_2)A_\nu^{(D)}(x_2)\right]^{\widetilde{}}-\bar{\psi}_l(x_2)\gamma^\nu\psi_l(x_2)A_\nu^{(D)}(x_2)\right\}:\ket{i}. \label{eq24}
\end{eqnarray}

This scattering process at tree level is described by the Feynman diagram given in Figure \ref{fig1}. The thermal propagator of the photon with momentum $\kappa$ is $\Delta_{\alpha\nu}(\kappa)=\Delta^{(0)}_{\alpha\nu}(\kappa)+\Delta^{(\beta)}_{\alpha\nu}(\kappa)$, where
\begin{eqnarray}
\Delta^{(0)}_{\alpha\nu}(\kappa)=\frac{\eta_{\alpha\nu}}{\kappa^2}\begin{pmatrix}
1 & 0\\0&-1
\end{pmatrix}\label{eq25}
\end{eqnarray}
is the temperature-independent part, and
\begin{eqnarray}
\Delta^{(\beta)}_{\alpha\nu}(\kappa)=-\frac{2\pi i\delta(\kappa^2)}{e^{\beta q_0}-1}\begin{pmatrix}
1 & e^{\beta q_0/2}\\e^{\beta q_0/2} & 1
\end{pmatrix}\eta_{\alpha\nu}\label{eq26}
\end{eqnarray}
is the temperature-dependent part of the propagator \cite{ale1}. The $\eta_{\alpha\nu}$ is the Minkowski metric with $\det{\eta}=-2$. Observe that, in the expression for the thermal propagator given in Eq. (\ref{eq26}) the chemical potential of the photon does not appear and this fact occurs because $\mu_{\text{photon}}=0$ as discussed earlier.

\begin{figure}[ht]
\centering
\includegraphics[scale=0.5]{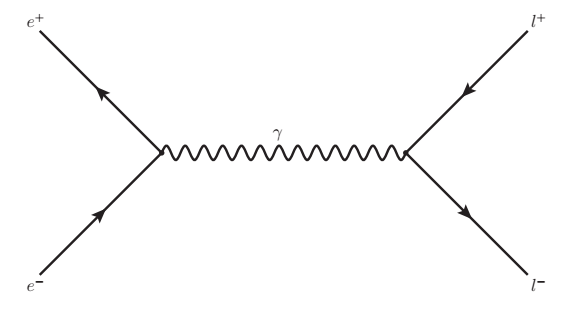}
\caption{Tree-level Feynman diagram of $e^{+}e^{-}\to l^{+}l^{-}$ scattering process.}
\label{fig1}
\end{figure}

At this point, from Eq. (\ref{eq24}), it is necessary to make a distinction between electronic and leptonic fields, since their masses are different. Looking at Eq. (\ref{eq07}), $\psi_e$ and $\psi_l$ are obtained by putting $M=m_e$ and $M=m_l$, respectively. Then
\begin{eqnarray}
\psi_e(x^\nu)=\int\frac{dp_xdp_z}{(2\pi)^2}\sum_{s,n}\sqrt{\frac{E_n+m_e}{2E_n}}\left[a_s(n,\vec{p}_{\cancel{y}})u_s(y,n,\vec{p}_{\cancel{y}})e^{-ip_\nu x^\nu_{\cancel{y}}}+b_s^{\dagger}(n,\vec{p}_{\cancel{y}})v_s(y,n,\vec{p}_{\cancel{y}})e^{ip_\nu x^\nu_{\cancel{y}}}\right]
\end{eqnarray}
for electrons, and
\begin{eqnarray}
\psi_l(x^\nu)=\int\frac{dp_xdp_z}{(2\pi)^2}\sum_{s,n}\sqrt{\frac{E_n+m_l}{2E_n}}\left[a_s(n,\vec{p}_{\cancel{y}})u_s(y,n,\vec{p}_{\cancel{y}})e^{-ip_\nu x^\nu_{\cancel{y}}}+b_s^{\dagger}(n,\vec{p}_{\cancel{y}})v_s(y,n,\vec{p}_{\cancel{y}})e^{ip_\nu x^\nu_{\cancel{y}}}\right]
\end{eqnarray}
for leptons. 

Note from Eqs. (\ref{eq25}) and (\ref{eq26}) that both equations do not depend on the Landau index $n$, and this is because the photon has no charge and therefore does not couple to the external magnetic field \cite{diracmag, dissertmag}.

\subsection{Limit of very strong magnetic fields}

Here a special case is addressed, a very strong magnetic field limit. This occurs when the amplitude $B$ is very large compared to $E^2-m_l^2$. In this regime, the only physical condition of interest is one in which the Landau index is zero ($n=0$), where the movement of the electron and lepton is restricted to a one-dimensional $z$-axis.

In this situation, the completeness relations given by Eqs. (\ref{eq27}) and (\ref{eq28}) take the form
\begin{eqnarray}
\sum_s u_s(y,\vec{p}_{\cancel{y}})\bar{u}_s(y^\prime,\vec{p}_{\cancel{y}})=\frac{1}{2(E_0+m_l)}\left[m_l(1-\Sigma_z)+\slashed{p}_{||}+\bar{\slashed{p}}_{||}\gamma^5\right]I_0(\xi^{+})I_0(\xi^{+\prime}),\label{eq56}
\end{eqnarray}
for the positive energy part and
\begin{eqnarray}
\sum_s v_s(y,\vec{p}_{\cancel{y}})\bar{v}_s(y^\prime,\vec{p}_{\cancel{y}})=-\frac{1}{2(E_0+m_l)}\left[m_l(1-\Sigma_z)+\slashed{p}_{||}+\bar{\slashed{p}}_{||}\gamma^5\right]I_0(\xi^{-})I_0(\xi^{-\prime}),\label{eq57}
\end{eqnarray}
for the negative energy part. Here have been used $\Sigma_z=i\gamma^1\gamma^2$, $\slashed{p}_{||}=\gamma^0p_0-\gamma^3p_3$, $\bar{\slashed{p}_{||}}=\gamma^0p_3-\gamma^3p_0$ and $\gamma^5=i\gamma^0\gamma^1\gamma^2\gamma^3$.

Note that $\mathcal{M}_{fi}$ is an average over thermal vacuum state. Then it is necessary to make the connection between finite and zero-temperature operators. The transition amplitude can be written using the Bogoliubov transformations Eqs. (\ref{eq08}) and (\ref{eq09}) and the tilde conjugation rules Eq. (\ref{eq55}). Thus, Eq. (\ref{eq24}) becomes
\begin{eqnarray}
\mathcal{M}_{fi}&=&-e^2F(\beta)\left(E_0+m_e\right)\left(E_0+m_l\right)\int d^4x_1d^4x_2\left[\bar{v}_{\lambda}(y_1,\vec{k}_{\cancel{y}})\gamma^\alpha u_s(y_1,\vec{p}_{\cancel{y}})e^{-i(p+k)\cdot x_{1\cancel{y}}}\right. \nonumber\\&\times &\left.
\bar{u}_{s^{\prime}}(y_2,\vec{p^{\prime}}_{\cancel{y}})\gamma^\nu v_{\lambda^\prime}(y_2,\vec{k^\prime}_{\cancel{y}})e^{i(p^{\prime}+k^{\prime})\cdot x_{2\cancel{y}}}\Delta_{\alpha\nu}(x_1-x_2)\right], \label{eq29}
\end{eqnarray}
where $F(\beta)$ is a thermal function that associates the Fermi-Dirac distribution of all particles in the following functional form 
\begin{eqnarray}
F(\beta)=\left[f(E_{e^{-}},\mu_e)+f(E_{e^{+}},-\mu_e)-1\right]\left[f(E_{l^{-}},\mu_l)+f(E_{l^{+}},-\mu_l)-1\right]\label{eq32}
\end{eqnarray}
whose analytical appearance is very dependent on the choice of the reference frame. In the center of mass we have
\begin{eqnarray}
p=\left(E,\vec{p}\right),\quad\quad k=\left(E,-\vec{p}\right),\quad\quad p^\prime=\left(E,\vec{p^\prime}\right),\quad\quad k^{\prime}=\left(E,-\vec{p^\prime}\right),
\end{eqnarray}
such that $m_e^2=E^2-|\vec{p}|^2$ and $m_l^2=E^2-|\vec{p^\prime}|^2$. Thus, the thermal function Eq. (\ref{eq32}) becomes
\begin{eqnarray}
F(\beta)=\frac{\sinh^2{\beta E}}{\left(\cosh{\beta\mu_e}+\cosh{\beta E}\right)\left(\cosh{\beta\mu_l}+\cosh{\beta E}\right)}.\label{eq58}
\end{eqnarray}

The photon propagator $\Delta_{\alpha\nu}(x_1-x_2)$ in the coordinate-space is given as
\begin{eqnarray}
\Delta_{\alpha\nu}(x_1-x_2)=i\int \frac{d^4\kappa}{(2\pi)^4} e^{-i\kappa\cdot(x_2-x_1)}\Delta_{\alpha\nu}(\kappa).
\end{eqnarray}

Performing the integration on $x_1$ and $x_2$ in Eq. (\ref{eq29}),  we get
\begin{eqnarray}
\mathcal{M}_{fi}&=&-ie^2F(\beta)\left(E_0+m_e\right)\left(E_0+m_l\right)\int\frac{d^4\kappa}{2\pi}\int dy_1dy_2\left[e^{i\kappa_yy_1}\bar{v}_{\lambda}(y_1,\vec{k}_{\cancel{y}})\gamma^\alpha u_s(y_1,\vec{p}_{\cancel{y}})\right.\nonumber\\&\times &\left.\delta^4_{\cancel{y}}(p+k-\kappa)
e^{-i\kappa_yy_2}\bar{u}_{s^{\prime}}(y_2,\vec{p^{\prime}}_{\cancel{y}})\gamma^\nu v_{\lambda^\prime}(y_2,\vec{k^\prime}_{\cancel{y}})\delta^4_{\cancel{y}}(p^\prime+k^\prime-\kappa)\Delta_{\alpha\nu}(\kappa)\right],\label{eq31}
\end{eqnarray}
where the definition of the Dirac delta function
\begin{eqnarray}
\int\frac{d^4x}{(2\pi)^4}e^{\pm(a+b)\cdot x_{\cancel{y}}}=\int \frac{dy}{2\pi}\delta^4_{\cancel{y}}(a+b) 
\end{eqnarray}
has been used.

Looking at the integration in $\kappa$, the Eq. (\ref{eq31}) becomes
\begin{eqnarray}
\mathcal{M}_{fi}&=&-ie^2F(\beta)\left(E_0+m_e\right)\left(E_0+m_l\right)\int dy_1dy_2\left[\delta(y_1-y_2)\bar{v}_{\lambda}(y_1,\vec{k}_{\cancel{y}})\gamma^\alpha u_s(y_1,\vec{p}_{\cancel{y}})\right.\nonumber\\&\times &\left.\bar{u}_{s^{\prime}}(y_2,\vec{p^{\prime}}_{\cancel{y}})\gamma^\nu v_{\lambda^\prime}(y_2,\vec{k^\prime}_{\cancel{y}})\delta^4_{\cancel{y}}(p^\prime+k^\prime-\kappa)\Delta_{\alpha\nu}(p+k)\right].
\end{eqnarray}
Since the delta function imposes conservation of the total momentum, the last equation can be written as
\begin{eqnarray}
\mathcal{M}_{fi}&=&-ie^2F(\beta)\left(E_0+m_e\right)\left(E_0+m_l\right)\nonumber
\\&\times&\int dy\left[\bar{v}_{\lambda}(y,\vec{k}_{\cancel{y}})\gamma^\alpha u_s(y,\vec{p}_{\cancel{y}})\Delta_{\alpha\nu}(p+k)\bar{u}_{s^{\prime}}(y,\vec{p^{\prime}}_{\cancel{y}})\gamma^\nu v_{\lambda^\prime}(y,\vec{k^\prime}_{\cancel{y}})\right].
\end{eqnarray}

For an initial unpolarized beam, the final and initial spin states are averaged. Then
\begin{eqnarray}
\langle|\mathcal{M}_{fi}|^2\rangle=\frac{1}{4}\sum_{\substack{s,s^\prime\\\lambda,\lambda^\prime}}|\mathcal{M}_{fi}|^2.
\end{eqnarray}
Using Eqs. (\ref{eq56}) and (\ref{eq57}) we obtain
\begin{eqnarray}
\langle|\mathcal{M}_{fi}|^2\rangle =2e^4H(\beta)\left(\frac{eB}{2\pi}\right) \left[m_e^2 m_l^2+\left(\text{E}^2+|p|^2\right) \left(\text{E}^2+|p^\prime|^2 \cos ^2\theta \right)\right],\label{eq35}
\end{eqnarray}
with
\begin{eqnarray}
H(\beta)=F^2(\beta)\left[\frac{1}{s^2}+\frac{(2\pi)^2\delta^2(s)}{(e^{
2\beta E}-1)^2}\right].\label{eq59}
\end{eqnarray}
Here has been used
\begin{eqnarray}
\int dy \left[I_0^2(\xi^{+})I_0^2(\xi^{-})\right]=\sqrt{\frac{eB}{2\pi}}.
\end{eqnarray}

Remember that we are dealing with the case where the background field is very strong, so this approach is considered $p_x=0$, which implies $p=(E,0,0,p_z)$. In addition, the angle that appears in Eq. (\ref{eq35}) is defined between $p$ and $p^\prime$, and since the motion is one-dimensional, it can assume only two values: $\theta=0$ or $\theta=\pi$ \cite{tiwari}.

Finally, putting Eq. (\ref{eq35}) into Eq. (\ref{eq42}) the cross-section for the scattering process reads
\begin{eqnarray}
\sigma_0(\beta)=\frac{\langle|\mathcal{M}_{fi}|^2\rangle}{8p^\prime_z s\sqrt{s-4m_e^2}}=\frac{e^4(\pi eB)}{2(2\pi)^2p^\prime_z s\sqrt{s-4m_e^2}}H(\beta)\left[m_e^2 m_l^2+\left(\text{E}^2+|p|^2\right) \left(\text{E}^2+|p^\prime|^2 \right)\right].\label{eq60}
\end{eqnarray}

When temperatures are low the function $F(\beta)$ becomes a step function around $E=\mu_0$, that is,
\begin{eqnarray}
\lim_{\beta\to\infty}F(\beta)=\begin{cases}0\quad\text{if}\quad E<\mu_0\\
 1\quad\text{if}\quad E>\mu_0\end{cases}
\end{eqnarray}
which leads to
\begin{eqnarray}
\lim_{\beta\to\infty}H(\beta)=\frac{1}{s^2}.
\end{eqnarray}
It is important to note that the cross-section vanishes for $E<\mu_0$ and becomes
\begin{eqnarray}
\sigma_0(T=0)=\frac{e^4(\pi eB)}{2(2\pi)^2p^\prime_z s^3\sqrt{s-4m_e^2}}\left[m_e^2 m_l^2+\left(\text{E}^2+|p|^2\right) \left(\text{E}^2+|p^\prime|^2 \right)\right],
\end{eqnarray}
for $E>\mu_0$. Or, more elegantly, as 
\begin{eqnarray}
\sigma_0(T=0)=\begin{cases}\frac{(\pi e B)\alpha^2}{s^3}\frac{\left[s^2+8m_e^2m_l^2-s\left(m_e^2+m_l^2\right)\right]}{\sqrt{(s-4m_e^2)(s-4m_l^2)}}\quad\quad &\text{if}\quad\quad E>\mu_0\\
0 \quad\quad &\text{if} \quad\quad E<\mu_0,\end{cases}\label{eq64}
\end{eqnarray}
where $\alpha=e^2/4\pi$ is the fine-structure constant. This result is the same obtained by \cite{proposta}, without the  EDM term (dipole term) that comes from the Lorentz violation, when the high energy approximation ($m_e=m_l=0$) is taken. 

The usual QED cross-section for the $e^{+}e^{-}\to l^{+}l^{-}$ scattering process is $\sigma_{\text{QED}}=4\pi\alpha^2/3s$ \cite{peskin}. In Figure \ref{fig7} is shown how $\sigma_0(T=0)$ behaves in terms of $\sigma_{\text{QED}}$ for $E>\mu_0$. Initially, it can be seen that there is an asymptotic limit at $s=4m_l^2$ and the cross-section obtained here exhibits an energy dependence $s^{-2}$ while the usual QED theory does for $s^{-1}$. In addition, due to the type of dependence of the $eB$ factor, stronger magnetic fields produce a linear increase in cross-section value.

\begin{figure}[ht]
\includegraphics[scale=0.7]{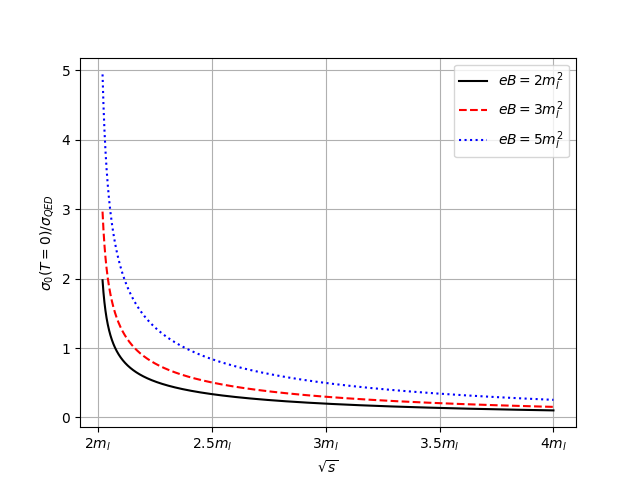}
\caption{The ratio cross-section at zero temperature and $\sigma_{\text{QED}}$ in terms of the centre of mass energy ($\sqrt{s}$) for the magnetic field amplitudes: $B=2m_l^2/e$ (black solid line); $B=3m_l^2/e$ (red dashed line) and $B=5m_l^2/e$ (blue dotted line). }\label{fig7}
\end{figure}

When the temperatures are very high, we obtain from Eqs. (\ref{eq58}) and (\ref{eq59}) the following approximation
\begin{eqnarray}
H(\beta)=\frac{(\beta E)^2}{4s^2}+\frac{(2\pi)^2\delta^2(s)}{8}\left\{\frac{1}{2}-\beta E-\left[\left(\frac{\mu_l}{E}\right)^2+\left(\frac{\mu_e}{E}\right)^2-6\right]\frac{(\beta E)^2}{8}\right\}
\end{eqnarray}
for any $E$ value. Then, the differential cross-section at this regime becomes 
\begin{eqnarray}
\lim_{T\to\infty} \sigma_0(\beta)= \frac{e^4(\pi eB)}{2(2\pi)^2p^\prime_z s\sqrt{s-4m_e^2}}\frac{(2\pi)^2\delta^2(s)}{16}\left[m_e^2 m_l^2+\left(\text{E}^2+|p|^2\right) \left(\text{E}^2+|p^\prime|^2 \right)\right],
\end{eqnarray}
which, as expected, does not depend on the chemical potential. The square Dirac delta function that appears in the solutions can be treated and assumes a real value considering its regularized forms \cite{deltaregular}.

Returning to Eq. (\ref{eq58}), taking $\mu_e=\mu_l=0$ limit, we get
\begin{eqnarray}
F(\beta)=\frac{\sinh^2{\beta E}}{\left(1+\cosh{\beta E}\right)^2}=\tanh^2{\frac{\beta E}{2}},\label{eq63}
\end{eqnarray}
and, in this way, the result presented in Eq. (\ref{eq60}) using Eq. (\ref{eq63}) is very similar to those obtained by \cite{praseletron}.

It is important to note that the weak magnetic field limit cannot be made by doing $B\to0$, and this fact comes from the strong field regime ($n=0$). To obtain the threshold of weaker background fields is necessary to follow the steps present in Eq. (\ref{eq61}). In other words, to make the limit of $B\to0$ it is necessary to take into account the continuous sum over all $n$-states, which is completely contrary to what we used initially, the very high field strength.

\begin{figure}[h]
\subfigure[aa1][The thermal function $F(\beta)$ in terms of inverse temperature for $\mu\neq0$ case (black solid line) given by Eq. (\ref{eq58}) and for $\mu=0$ case (red dashed line) given by Eq. (\ref{eq63}) in $E=2\mu$ environment.]{\includegraphics[width=8.1cm]{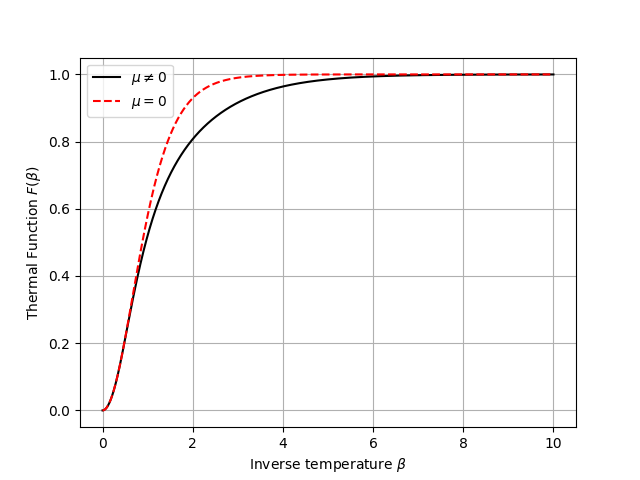}\label{fig6a}}
\subfigure[bb1][The thermal function $F(\beta)$ in terms of inverse temperature for $\mu\neq0$ case (black solid line) given by Eq. (\ref{eq58}) and for $\mu=0$ case (red dashed line) given by Eq. (\ref{eq63}) in $E=\mu/2$ environment.]{\includegraphics[width=8.1cm]{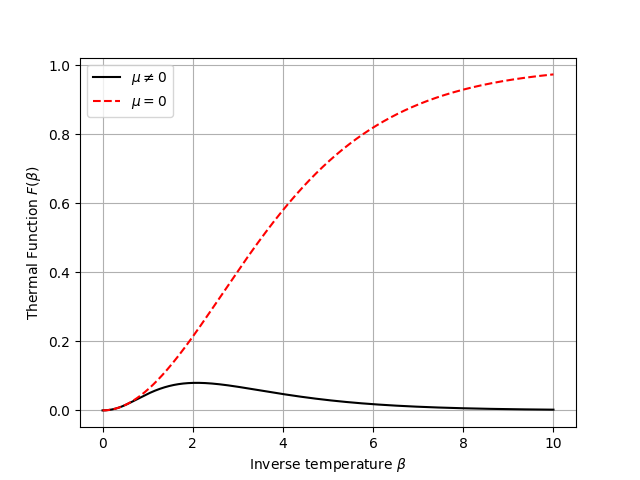}\label{fig6b}}

\caption{The thermal functions in terms of the inverse temperature.\label{fig6}}
\end{figure}

Regarding the thermal functions Eqs. (\ref{eq58}) and (\ref{eq63}), Figure \ref{fig6} shows a comparison between them. Note that, in Figure \ref{fig6a}, a system with energy $E>\mu$ shows similar behavior for both functions, although the presence of the chemical potential changes some regimes by an amount. In addition, the limits for high or lower temperatures are the same.  

Now, looking at Figure \ref{fig6b} it can be seen that, for energies less than $\mu$, the reaction cannot occur except for some short regime. But, as discussed, for very high temperatures, or low $\beta$-values, the interpretation of the chemical potential as the energy required to add a particle to the system does not even make sense.

It is important to note that the masses of other leptons are larger than those of electrons; hence, the effect of an external field on an electron is more pronounced than on other leptons. Consequently, the field magnitudes that need to be taken into account are those that can be felt by all leptons. Furthermore, these magnitudes are so intense that the $n=0$ approximation can be applied. In simpler terms, when referring to a very strong field, we mean magnitudes so large that they affect all leptons, constraining them to motion only in the $z-$direction. In this regime, the energy and chemical potential of all particles do not depend on $B$, but only the spinors do. It is due to this dependence that the final cross-section is directly influenced by the external field.

Finally, the thermal function Eq. (\ref{eq58}) is strongly dependent on the lepton flavor, since $\mu$ for $T=0$ lies around the particle mass. For reactions where the final products are muon or tau fermions, one can take $\mu_e=0$ since these masses are many times greater than that of the electron. For the Bhabha scattering $e^{+}e^{-}\to e^{+}e^{-}$, however, this assumption needs a little bit more attention, since that $m_l=m_e$ and, further, it is necessary to calculate another additional diagram in addition to Figure \ref{fig1}, which will not be done in this paper.

\section{Conclusions}\label{secconclusion}

An electron-positron pair decaying into lepton-antileptons, subject to thermal effects and a strong background classical magnetic field, has been presented. The effect due to an external field is inserted into the problem through a redefinition of the fermionic field operator that includes a dependence on $y-$coordinate and Landau level $n$. It is because in the quantization process, instead of writing the wave function as a well-localized combination of plane waves with defined momenta, the $y-$component is present in the spinor expression, leading to a $n-$dependence of the energy relation. Furthermore, the energy function comes from a new dispersion relation and the cross-section is calculated, where both quantities involve these levels. These modifications are discussed. Since the order of the external field in cosmological situations is very strong, the assumption $n_{\text{max}}=0$ is valid and this guarantees our choice of approach.

Temperature and chemical potential are introduced into the theory using the TFD formalism. A functional relation between cross-section and temperature, as well as for the chemical potential, is obtained. Although most published papers do not consider the quantity $\mu$, we can see that it makes a considerable difference in the final result. More explicitly, the chemical potential at zero temperature determines when the reaction can or cannot occur. Under these conditions, the reaction is analyzed for the very strong magnetic field limit. Limits and estimates for temperature and chemical potential effects are discussed. A temperature-dependent chemical potential has been obtained. It is important to note that the cross-section at finite temperature is not directly measurable in any many-body process. Such a quantity, in this context, is an important input to treat a gas at low density through a kinetic theory approach such as a Boltzmann equation. Then the finite temperature cross-sections will form part of the collision term that would lead to a distribution function by a solution of the Boltzmann equation. Furthermore, the results obtained here are important for a better understanding of the early Universe at finite temperature and density.

\section*{Acknowledgments}

This work by A. F. S. is partially supported by National Council for Scientific and Technological Development - CNPq project No. 313400/2020-2. D. S. C. thanks CAPES for financial support.

\section*{Data Availability Statement}

No Data associated in the manuscript.

%%%%%%%%%%%%%%%%%%%%%%%%%%%%%%%%%%%%%%%%%%%%%%%%%%%%%%%%%%%%%%%%%%%%%%%%%%%%%%%%%%%%%%%%%%%%%%%%%%%%%%%%%%%%%%%%%

\global\long\def\link#1#2{\href{http://eudml.org/#1}{#2}}
 \global\long\def\doi#1#2{\href{http://dx.doi.org/#1}{#2}}
 \global\long\def\arXiv#1#2{\href{http://arxiv.org/abs/#1}{arXiv:#1 [#2]}}
 \global\long\def\arXivOld#1{\href{http://arxiv.org/abs/#1}{arXiv:#1}}

%%%%%%%%%%%%%%%%%%%%%%%%%%%%%%%%%%%%%%%%%%%%%%%%%%%%%%%%%%%%%%%%%%%%%%%%%%%%%%%%%%%%%%%%%%%%%%%%%%%%%%%%%%%

\end{document}